# Gap opening of single-layer graphene under the continuum model

**Xin Lin, Hailong Wang, Hui Pan and Huaizhe Xu\***

*State Key Laboratory of Software Development Environment, and Department of Physics, Beihang University, Beijing 100191, P. R. China*

*Abstract:* Gap opening at the Dirac point of the single-layer graphene with periodic scalar and vector potentials has been theoretically investigated under the continuum model. The symmetry analysis indicates that the two-fold degeneracy at the Dirac point can be lifted when the potentials break both the chiral symmetry and the time-reversal symmetry. A gap equation at the Dirac point is obtained analytically with perturbation theory. It is shown that a mass term at the Dirac point would be generated by coupling of vector and scalar potentials. This gap equation could be considered as a criterion for gap opening at the Dirac point, which is confirmed by the numerical calculation. Furthermore, the bandgap from the gap equation agrees well with the exact result, when the applied potentials are weak.

------------------------------------

\*To whom all correspondence should be sent: hzxu@buaa.edu.cn (Huaizhe Xu);

Tel. & Fax: +86-010-82317935.

# 1. Introduction

Graphene [1], a monolayer of carbon atoms with honeycomb lattice structure, has attracted a wealth of studies in both fundamental physics and potential applications in information technology [2, 3]. Unlike normal semiconductor, the single-layer graphene is a semimetal in which the valence and conduction bands touch at the two Dirac points [4, 5]. Its low-energy excitation quasi-particles present a linear energy dispersion behavior as massless Dirac fermions [6]. Graphene electrons could penetrate high and wide electric barrier without reflection, known as Klein tunneling [7]. Therefore, the pristine graphene could not be used to make a field effect transistor [8, 9] due to its gapless band structure. Thus, opening a bandgap in graphene is especially essential for the application of electronic devices.

The physical process in graphene can be divided into two classes depending on whether couple the two Dirac points or not [5, 10]. (1) The tight-binding model: it is described by the nearest-neighbor hopping on honeycomb lattice, which considers the intervalley scattering between K and K', breaking the symplectic time-reversal symmetry [11]. (2) The continuum model: it is a well model when we consider a fabrication superlattice is applied to the graphene sheet. Experimentally, the size of the applied potential is much larger than the graphene lattice constant, so that the intervalley scattering is neglected, and the charge carrier is described by the single valley Hamiltonian.

In the tight-binding model, the gapless band structure of the Dirac point is protected by the inversion symmetry P and the orthogonal time-reversal symmetry T [11]. A bandgap could be generated at the Dirac point through breaking the P symmetry by graphene-substrate interaction [12], patterned hydrogen adsorption [13], controlled adatom deposition [14], physical cutting single-layer graphene into nanoribbons[15, 16] or etching it into antidot lattices [17]. Alternatively, the degeneracy at Dirac point could be lifted by breaking T symmetry, as demonstrated by Haldane's model [18].

Recently, the electronic properties of single-layer graphene under the continuum model are explored [19-24]. With the single valley Hamiltonian, several schemes have been proposed to open a bandgap at the Dirac point: Ref. 25 has shown that a bandgap could be generated at the Dirac point by overlapping of periodic electric field and magnetic field [25]; Ref. 26 have proposed that a bandgap at Dirac point could be generated and tuned by using a noncentrosymmetric two-dimensional electric superlattice [26]. However, a general scheme for opening the bandgap at the Dirac point by periodic scalar and vector potentials is absent, and the relationships between the degeneracy of the Dirac point and the symmetry of the Hamiltonian are not established.

In this paper, we investigate the scheme of gap opening in the single-layer graphene under the continuum model. The two-fold degeneracy at the Dirac point is analyzed by symmetry argument. Furthermore, a gap equation for lifting the two-fold degeneracy is analytically derived by a two-fold degeneracy perturbation method. It is shown that a bandgap could be generated at the Dirac point with combined scalar and vector potentials. Finally, we propose an example with opening gap at the Dirac point by the gap equation, and the results from the perturbation method are compared with the exact results calculated by linear combination of plane wavefunctions method. This gap equation will provide deep insights on the bandgap engineering of the single-layer graphene.

This paper is organized as follows. In Sec. 2, we discuss the two-fold degeneracy at the Dirac point by analyzing the symmetry of the single valley Hamiltonian. In Sec. 3, we introduce the two-fold degeneracy perturbation method and derive the gap equation for gap opening at the Dirac point. In Sec. 4, an example with opening gap at the Dirac point is proposed, and the perturbation result is compared with the exact result. We summarize our results and draw conclusions in sec. 5.

## 2. Symmetry of single valley Hamiltonian

The single valley Dirac-like Hamiltonian with periodic scalar and vector potentials could be expressed as (with $\hbar = c = 1$):

$$H = v_F\left(-i\nabla + e\vec{A}\right)\cdot \sigma + eU \cdot I, \tag{1}$$

where $v_F = 10^6 \, m/s$ is the Fermi velocity, $e$ is the absolute value of the electron charge, $U$ is the scalar potential, $\vec{A}$ is the vector potential, $\sigma = (\sigma_x, \sigma_y)$ is a 2×2 Pauli matrices vector, and I is the 2×2 unit matrix.

We assume that the strength of both electric field $U$ and magnetic field $B$ are zero on average, that is <U(x, y)>=0, <B(x, y)>=0. The scalar potential $\vec{A}$ could be obtained by:

$$\nabla \times \vec{A} = B\hat{z}, \tag{2a}$$

where $\vec{A} = (A_x, \ A_y, \ 0)$. By choosing the Coulomb gauge:

$$\nabla \cdot \vec{A} = 0, \tag{2b}$$

we get <$A_x$(x, y)>=0, <$A_y$(x, y)>=0. (see Appendix Eq. (A.8))

The total Hamiltonian could be rewritten as:

$$H = H_0 + H', \tag{3a}$$

with

$$H_0 = v_F \begin{pmatrix} 0 & \hat{k}_x - i\hat{k}_y \\ \hat{k}_x + i\hat{k}_y & 0 \end{pmatrix}, \tag{3b}$$

and

$$H' = \begin{pmatrix} eU & ev_F(A_x - iA_y) \\ ev_F(A_x + iA_y) & eU \end{pmatrix}. \tag{3c}$$

Here, $H'$ is the intrinsic single valley Hamiltonian, and $H'$ is the Hamiltonian induced by the external potentials.

It is noted that $H_0$ holds the time-reversal symmetry:

$$H_0 = SH_0 S^{-1} = \sigma_y H_0^* \sigma_y, \tag{4}$$

and the chiral symmetry:

$$-H_0 = \sigma_z H_0 \sigma_z. \tag{5}$$

Here, $S = \begin{pmatrix} 0 & 1 \\ -1 & 0 \end{pmatrix}.c$, with $c$ the complex conjugation operator, and $\sigma_z$ is the Pauli matrix.

With pure periodic scalar potential, $H' = eU \cdot I$, the chiral symmetry is broken, while the total Hamiltonian still hold the time-reversal symmetry:

$$SHS^{-1} = H. \tag{6}$$

For any Bloch state $\psi_k = \begin{pmatrix} \phi_{ka} \\ \phi_{kb} \end{pmatrix}.e^{i\bar{k}\cdot\bar{r}}$, there is always a corresponding state $S\psi_k = \begin{pmatrix} c\phi_{kb} \\ -c\phi_{ka} \end{pmatrix}.e^{-i\bar{k}\cdot\bar{r}}$. $\psi_k$ and $S\psi_k$ are orthogonal states, which could be labeled as (E, **k**) and (E, -**k**). These two states degenerate at **k**=0 point.

With pure periodic vector potential, $H' = v_F e\vec{A} \cdot \sigma$, the time-reversal symmetry is broken, while the chiral symmetry of the system is held:

$$-H_0 = \sigma_z H_0 \sigma_z. \tag{7}$$

Therefore, if $\psi_E$ is an eigenstate of the energy E, $\sigma_z \psi_E$ is the eigenstate of energy –E. It has been demonstrated that the E=0 mode always exists in this system [25, 27]. Due to the chiral symmetry, E=0 mode would be two-fold degeneracy.

It is obvious that the two-fold degeneracy at the Dirac point will never be lifted by scalar potential or vector potential, separately. In order to lift the two-fold degeneracy at the Dirac point, the time-reversal symmetry and the chiral symmetry should be broken at the same time. It is anticipated that the two-fold degeneracy could be lifted when both scalar and vector potentials are applied. This expectation will be confirmed by second-order perturbation calculation in the following section, and an analytic expression for opening a bandgap at the Dirac point will be also derived.

## 3. Analytical expression by second-order perturbation method

To analyze the two-fold degeneracy at the Dirac point, we treat the weak scalar and vector potentials perturbatively. As $<A_x(x, y)>=0$, $<A_y(x, y)>=0$, and $<U(x, y)>=0$, the leading-order perturbation is obviously zero. Then, the secular equation of the second-order perturbation is obtained as:

$$\begin{pmatrix} H_{11} & H_{12} \\ H_{21} & H_{22} \end{pmatrix} \psi - E*I\psi = 0, \tag{8}$$

with

$$H_{ij} = \sum_{s,k} \frac{\langle \psi_i | H' | \psi^0_{s,k} \rangle \langle \psi^0_{s,k} | H' | \psi_j \rangle}{0 - E}, \tag{9}$$

where $\psi^0_{s,k}$ is the eigenstate of $H_0$, and $\psi_i$ is the zero-order wavefunctions (i=1, 2) at the Dirac point that is two-fold degeneracy. The box normalization wavefunctions are given as:

$$\psi^0_{s,k}(x, y) = \frac{1}{\sqrt{L^2}} \cdot \frac{1}{\sqrt{2}} \begin{pmatrix} 1 \\ s.e^{i\theta_k} \end{pmatrix} e^{ik_x x}.e^{ik_y y}, \tag{10}$$

$$\psi_1 = \frac{1}{\sqrt{L^2}} \begin{pmatrix} 1 \\ 0 \end{pmatrix}, \quad \psi_2 = \frac{1}{\sqrt{L^2}} \begin{pmatrix} 0 \\ 1 \end{pmatrix}, \tag{11}$$

where $s= \pm 1$ is the band index, $\frac{1}{\sqrt{L^2}}$ is the box normalization factor, $\mathbf{k}=(k_x, k_y)$ is the wave vector of the plane wavefunction and $\theta_k$ is the polar angle of the wave vector $\mathbf{k}$.

By inserting Eq.(10) and Eq.(11) into Eq.(9), we get [28]

$$H_{11} = \int_{s=1,k_x>0,k_y>0.} dE.kd\theta_k \rho(E,\theta_k) I_{11}. \tag{12}$$

Here,

$$\rho(E,\theta_k) = \frac{L^2 \cdot E}{2\pi \cdot v_F^2} \cdot \frac{1}{2\pi k} = \frac{L^2}{4\pi^2 v_F^2}. \tag{13}$$

$$I_{11} = \frac{1}{2}\frac{1}{L^4}\frac{1}{E}e^2 v_F 8\cos\theta_k \left( \iint_{L^2} dxdy\, U\cos(k_x x + k_y y) \iint_{L^2} dxdy\, A_y \sin(k_x x + k_y y) - \iint_{L^2} dxdy\, U\sin(k_x x + k_y y) \iint_{L^2} dxdy\, A_y \cos(k_x x + k_y y) \right)$$

$$+ \frac{1}{2}\frac{1}{L^4}\frac{1}{E}e^2 v_F 8\cos\theta_k \left( \iint_{L^2} dxdy\, U\cos(k_x x - k_y y) \iint_{L^2} dxdy\, A_y \sin(k_x x - k_y y) - \iint_{L^2} dxdy\, U\sin(k_x x - k_y y) \iint_{L^2} dxdy\, A_y \cos(k_x x - k_y y) \right)$$

$$\tag{14}$$

with $k_x = k\cos\theta_k$, $k_y = k\sin\theta_k$, and $k = E/v_F$.

From Eq. (9) to Eq. (14), we can further get $H_{22}= -H_{11}$, $H_{12}= H_{21}=0$. As a result, the eigenvalues of the secular equation of Eq. (8) is:

$$E = \pm H_{11}. \tag{15}$$

It is obvious that when

$$H_{11} \neq 0, \tag{16}$$

the two-fold degeneracy at the Dirac point will be lifted and a bandgap width of $2H_{11}$ will be generated. Thus, Eq. (16) is a gap equation at the Dirac point.

$H_{11}$ is a function of the coupling between scalar potential $U$ and vector potential $A_y$, and it could be regarded as a mass term of Dirac fermions [25]. When U(x,y)=0 or $A_y$(x,y)=0, we always have $H_{11}$=0, which means that the two-fold degeneracy at the Dirac point always hold with scalar potential or vector potential, separately. When U(x,y) $\neq$0 and $A_y$(x,y) $\neq$0, we normally get $H_{11}\neq$0. However, we may get $H_{11}$=0 for some specific spatial configuration of scalar and vector potentials, which correspond to the accidental degeneracy, such as $U(x,y) = U(-x,-y)$ and $A_y(x,y) = A_y(-x,-y)$, or $U(x,y) = -U(-x,-y)$ and $A_y(x,y) = -A_y(-x,-y)$.

The gap equation obtained by the second-order perturbation method is consistent with above symmetry analysis. When both scalar and vector potentials are applied, the two-fold degeneracy will be lifted and a bandgap will be generated at the Dirac point, except in some accidental degeneracy cases. With pure scalar or vector potential, the two-fold degeneracy always holds and the single graphene remains gapless at the Dirac point.

It is worthy to note that our discussion about the gap opening above is also applicable to the other Dirac spectrum system, such as the surface state of the 3D strong topological insulator [30], since their electron excitations are described by a similar 2×2 Weyl Hamiltonian.

## 4. Numerical examples

To confirm the gap opening of single graphene by applying the scalar and vector potentials, we propose a model structure based on the gap equation. For simplification, we take a square lattice of the scalar and vector potentials, as shown in Fig. 1:

$$U(x,y)=U_1\left(\theta_{(x_0,0),0.25}(x,y)-\theta_{(0.5+x_0,0),0.25}(x,y)\right), \quad B(x,y)=B_1\cos 2\pi(x+y). \quad (17)$$

Where $\Theta$ is a function defined as: when $|\vec{r}(x,y)-\vec{r}_o(x_0,y_0)|\leq 0.25$, $\Theta$ is 1, otherwise 0. The first index of function $\Theta$ could be understood as the axis of the circle centre, and the second index as the radius.

Figs. 2(a) and 2(b) show the renormalized energy band along the two high-symmetry directions Γ-M and Γ-X in superlattice Brillouin zone (SBZ). The energy band presents a bandgap when $x_0$=0, but gapless when $x_0$=0.25 at the Dirac point. The bandgap at the Dirac point as a function of $x_0$ in one period is shown in Fig. 2(c). It can be seen that the bandgap at the Dirac point changes from gapped to gapless periodically with $x_0$. A bandgap presents at the Dirac point for most spatial configurations of the scalar and vector potentials, except in some specific spatial configurations, such as $x_0$=0.25 or $x_0$=0.75. We have $A_y(x,y)=-A_y(-x,-y)$ and $U(x,y)=-U(-x,-y)$ when $x_0$=0.25 or $x_0$=0.75. As we have demonstrated in Sec. III, the coupling term $H_{11}$=0 and the bandgap would not be generated for these configurations. Here, the gap equation plays a role of criterion for gap opening at the Dirac point well.

By inserting Eq. (17) into Eq. (14), we get $H_{11}$=BUcos($2\pi \cdot x_0$)I, and the bandgap width at the Dirac point:

$$\Delta_{gap}=|2B.U.\cos(2\pi.x_0).I|. \quad (18)$$

Here, I is a definite integral and I=3.48×10$^{-3}$. The bandgap at the Dirac point calculated by perturbation method and the linear combination of plane wavefunctions method are compared in Figs. 2(d), 2(e) and 2(f). It is clear that these two results are consistent well with each other when the scalar and vector potentials are weak. Since the difference between these two results is less than 5 percent when 0<U <5.0 and 0<B<10, the gap equation could provide a quantitative result of the gap opening problem for the weak potentials. As we choose $L_x$=$L_y$=20nm, $B_0$ ~ 1.6 T, the effective magnetic strength for the purterbation method is from 0 to 16 T, sufficient to the highest magnetic strength that could be realized.

To further examine our symmetry argument and the gap equation, we analyze the possibility of gap opening in several models reported in the recent literatures.

Case 1: With pure periodic scalar potential or vector potential: <U(x, y)>=0, B(x, y)=0 (Refs. 19, 20 and 21); or <B(x, y)>=0, <$A_x$(x, y)>=0, <$A_y$(x, y)>=0 and U(x, y)=0 (Ref. 22, 23 and 24). As we have demonstrated, there would be no bandgap at the Dirac point, which is the same as the publications previously mentioned. When <U(x,y)>≠0, U(x) could be divided into two parts as U(x,y)=$U_1$(x,y)+$U_2$(x,y) with <$U_1$(x,y)>=0 and $U_2$(x,y) a constant, which only shift the whole band structure upward or downward. The part <U(x,y)>≠0 does not affect the gap opening at the Dirac point.

Ref. 26 claims that the two-fold degeneracy of the Dirac point would be split by applying a triangular superlattice or a square superlattice without inversion symmetry. It is controversial to our gap equation. The band structure of the second example in Ref. 26 is recalculated numerically by linear combination of plane wavefunctions method. As shown in Fig. 3, there is no bandgap at the Dirac point can be seen. It is also noted that the band structure in a similar configuration of the first example in Ref. 26 is studied by first principle, which gives zero energy gap at the Dirac point [31]. We thus believe that the bandgap presented in Ref. 26 may be due to a mistake in numerical calculations.

Case 2: Combined periodic scalar and vector potentials <B(x, y)>=0, <$A_x$(x, y)>=0, <$A_y$(x, y)>=0 and <U(x, y)>=0 (Refs. 25, 28 and 29). There would be a bandgap at the Dirac point except at some accidental degeneracy cases. In the first example of Ref. 25, the magnetic field and the electric fields are given as:

$$B(x, y) = B_2 (\sin 2\pi x + \sin 2\pi y), \quad U(x, y) = U_2 (\sin 2\pi x + \sin 2\pi y). \quad (19)$$

We get the vector potential $A_y$:

$$A_y(x, y) = -B_2 \frac{1}{2\pi} \cos 2\pi x. \quad (20)$$

In this configuration: $U(x, y) = -U(-x, -y)$ and $A_y(x, y) = A_y(-x, -y)$, we get $H_{11} \neq 0$, and a bandgap would be generated at the Dirac point, as shown clearly in Fig.4.

## 5. Conclusions

The gap opening at the Dirac point of the single-layer graphene has been theoretically investigated under the continuum model. The two-fold degeneracy at the Dirac point is protected by its chiral symmetry or the time-reversal symmetry. A bandgap could only be generated with combined scalar and vector potentials. We derive a gap equation for gap opening at the Dirac point by using a two-fold degeneracy perturbation method. The bandgap originates from the coupling between scalar and vector potentials, and behaves like a mass term of Dirac fermions. This gap equation is considered as a criterion for gap opening at the Dirac point, which is confirmed by our model and the other reported model structures. Furthermore, the bandgap obtained by numerical calculation agree exactly with the perturbation calculation, when the applied potentials are weak. Thus, we believe this gap equation could ultimately solve the gap opening problem with periodic scalar and vector potentials. The gap opening in single-layer graphene is of significant importance in design and fabrication of graphene-based electronics devices. The periodic scalar and vector potentials which are equivalent to electric and magnetic fields are very easy to produce and control, which is extensively applied in the present state-of-art semiconductor band engineering.

**Acknowledgement:**

This work is financially supported by State Key Laboratory of Software Development Environment under Grant No. SKLSDE-2011ZX-17, and National Science Foundation of China under Grant No. 10974011.

## Appendix

In this part, we derive the formula for the energy dispersion of the single-layer graphene under the periodic electric and magnetic potentials

$$U(r) = U(r + \vec{R}_{n',m'}),  \quad (A.1a)$$

$$B(r) = B(r + \vec{R}_{n',m'}),  \quad (A.1b)$$

where $\vec{R}_{n,m} = n' \cdot \vec{l}_1 + m' \cdot \vec{l}_2$, $\vec{l}_1$, $\vec{l}_2$ are the lattice constant of the periodic potentials, and $n'$, $m'$ are integer.

The electric field $U$ and the magnetic field $B$ could be expressed in Fourier expansion as:

$$U(\vec{r}) = \sum_{G_{n,m}} U(\vec{G}_{n,m}) e^{i\vec{G}_{n,m} \cdot \vec{r}},  \quad (A.2a)$$

$$B(\vec{r}) = \sum_{G_{n,m}} B(\vec{G}_{n,m}) e^{i\vec{G}_{n,m} \cdot \vec{r}},  \quad (A.2b)$$

with the Fourier component,

$$U(\vec{G}_{n,m}) = \frac{1}{L_0^2} \int U(\vec{r}) e^{-i\vec{G}_{n,m} \cdot \vec{r}} d\vec{r},  \quad (A.3a)$$

$$B(\vec{G}_{n,m}) = \frac{1}{L_0^2} \int B(\vec{r}) e^{-i\vec{G}_{n,m} \cdot \vec{r}} d\vec{r}.  \quad (A.3b)$$

Here, $G_{n,m} = nG_1 + mG_2$ is the reciprocal lattice. $L_0^2 = l_1 \cdot l_2$ is the area of a unit cell. The reciprocal lattice constant $G_1$ and $G_2$ could be obtained by the equations: $G_i \cdot l_j = 2\pi \cdot \delta_{i,j}$.

The scalar potential $A_x$ and $A_y$ could be also expressed in Fourier expansion as:

$$A_x(\vec{r}) = \sum_G A_x(\vec{G}_{n,m}) e^{i\vec{G}_{n,m} \cdot \vec{r}},  \quad (A.4a)$$

$$A_y(\vec{r}) = \sum_G A_y(\vec{G}_{n,m}) e^{i\vec{G}_{n,m} \cdot \vec{r}}.  \quad (A.4b)$$

From Eqs.(2a) and (2b), we arrive at,

$$\tfrac{\partial}{\partial x} A_y - \tfrac{\partial}{\partial y} A_x = B,  \quad (A.5a)$$

$$\tfrac{\partial}{\partial x} A_x + \tfrac{\partial}{\partial y} A_y = 0.  \quad (A.5b)$$

Combining Eq.(A1a) and (A1b) as $\tfrac{\partial}{\partial x}(A1a) + \tfrac{\partial}{\partial y}(A1b)$ and $\tfrac{\partial}{\partial y}(A1a) - \tfrac{\partial}{\partial x}(A1b)$, we can further arrive at,

$$\left(\tfrac{\partial^2}{\partial x^2} + \tfrac{\partial^2}{\partial y^2}\right) A_y = \tfrac{\partial}{\partial x} B,  \quad (A.6a)$$

$$\left(\frac{\partial^2}{\partial x^2}+\frac{\partial^2}{\partial y^2}\right)A_x = -\frac{\partial}{\partial y}B. \qquad (A.6b)$$

By inserting Eq.(A3) and (A4) into Eq.(A2), we get the Fourier component of $A_x$ and $A_y$:

$$A_y(\vec{G}_{n,m}) = -i\frac{G_{n,m_x}}{|\vec{G}_{n,m}|^2}B(\vec{G}_{n,m}), \qquad (A.7a)$$

$$A_x(\vec{G}_{n,m}) = i\frac{G_{n,m_y}}{|\vec{G}_{n,m}|^2}B(\vec{G}_{n,m}). \qquad (A.7b)$$

where $G_{n,m_x}$ and $G_{n,m_y}$ are the x-direction component and y-direction component of the wave vector $\vec{G}_{n,m}$. As <B(x, y)>=0, the Fourier component of the magnetic field $B(\vec{G}_{n,m}=0)=0$. Thus, the Fourier component of the scalar potential could be chosen as $A_x(\vec{G}_{n,m}=0)=0$ and $A_y(\vec{G}_{n,m}=0)=0$.

Finally, we get

$$<A_x(x,y)>=0, \quad <A_x(x,y)>=0. \qquad (A.8)$$

The Bloch wavefunction expressed in a linear combination of plane wavefunctions expansion:

$$\psi_k = \sum_{s,\vec{G}_{n,m}} C(s,\vec{G}_{n,m})\frac{1}{\sqrt{2}}\begin{pmatrix}1\\s.e^{i\theta_{\vec{G}_{n,m}+\vec{k}}}\end{pmatrix}.e^{i\vec{G}_{n,m}\cdot\vec{r}}.e^{i\vec{k}\cdot\vec{r}}, \qquad (A.9)$$

where $s=\pm 1$ is the band index, and $\vec{G}_{n,m}$ is the reciprocal lattice.

By inserting the Bloch wavefunction and the periodic potential into the Schrödinger equation:

$$(H_0 + H')\psi_k(\vec{r}) = E\psi_k(\vec{r}), \qquad (A.10)$$

We obtain the energy dispersion $E=f(\vec{k})$ by the following set of linear equations:

$$\sum_{s,\vec{G}} C(s,\vec{G})\sum_{\vec{G}_{n,m}} \frac{1}{2}\left(\left(1+ss'e^{i(\theta_{G+k}-\theta_{G'+k})}\right)U(\vec{G}_{n,m})+se^{i\theta_{G+k}}\left(A_x(\vec{G}_{n,m})-iA_y(\vec{G}_{n,m})\right)+s'e^{-i\theta_{G'+k}}\left(A_x(\vec{G}_{n,m})+iA_y(\vec{G}_{n,m})\right)\right)\delta_{G+G_{n,m},G'} = \left(E-s'v_F|\vec{G}'+\vec{k}|\right)C(s',\vec{G}')$$

$$(A.11)$$

For all of our numerical calculation, the number of the basis plane wavefunctions are chosen as 1250, that n is chosen as integer from -12 to 12, and m also integer from -12 to 12. The dimensionless units are chosen as: $l \to L_0 l$, $k \to k_0 k$, $E \to E_0 E$, $A \to B_0 L_0 A$, where $k_0 = 1/L_0$, $E_0 = \hbar v_F/L_0$, $B_0 = \hbar/eL_0^2$.

**Figures captions**

**Figure 1.** (a) Schematic of a periodic potential applied to graphene sheet. (b) The first SBZ, in which $\Gamma$-M and $\Gamma$-X are the high-symmetry direction across the $\Gamma$ point.

**Figure 2.** The band structure of single-layer graphene along $\Gamma$-M direction (solid line), and along the $\Gamma$-X direction (dashed line) of the example, in which $U_1=5.0$, $B_1=\sqrt{2} \cdot 2\pi$, $x_0=0$ (a), and $x_0=0.25$ (b). The bandgap at the Dirac point versus $x_0$ when $U_1=5.0$, and $B_1=\sqrt{2} \cdot 2\pi$ (c). The bandgap at the Dirac point vs. U when $B=\sqrt{2} \cdot 2\pi$, and $x_0=0$ (d); vs. B when $U=5.0$, and $x_0=0$ (e); vs. $x_0$ when $U=5.0$, and $B=\sqrt{2} \cdot 2\pi$ (f).

**Figure 3.** The band structure of the second example of Ref. 26 along $\Gamma$-M direction (solid line), and along the $\Gamma$-X direction (dashed line). All of the parameters are the same as those in Ref. 26.

**Figure 4.** The band structure of the first example of Ref. 25 along $\Gamma$-M direction (solid line), and along the $\Gamma$-X direction (dashed line) with $U_2=1.0$, and $B_2=2\pi$.

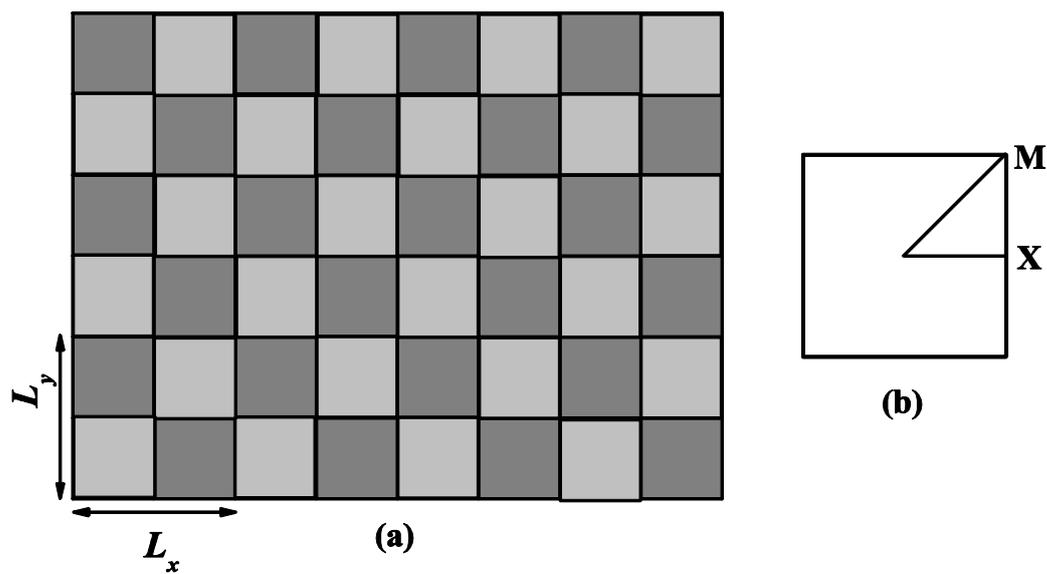

*Fig.1 by Xin Lin et al.*

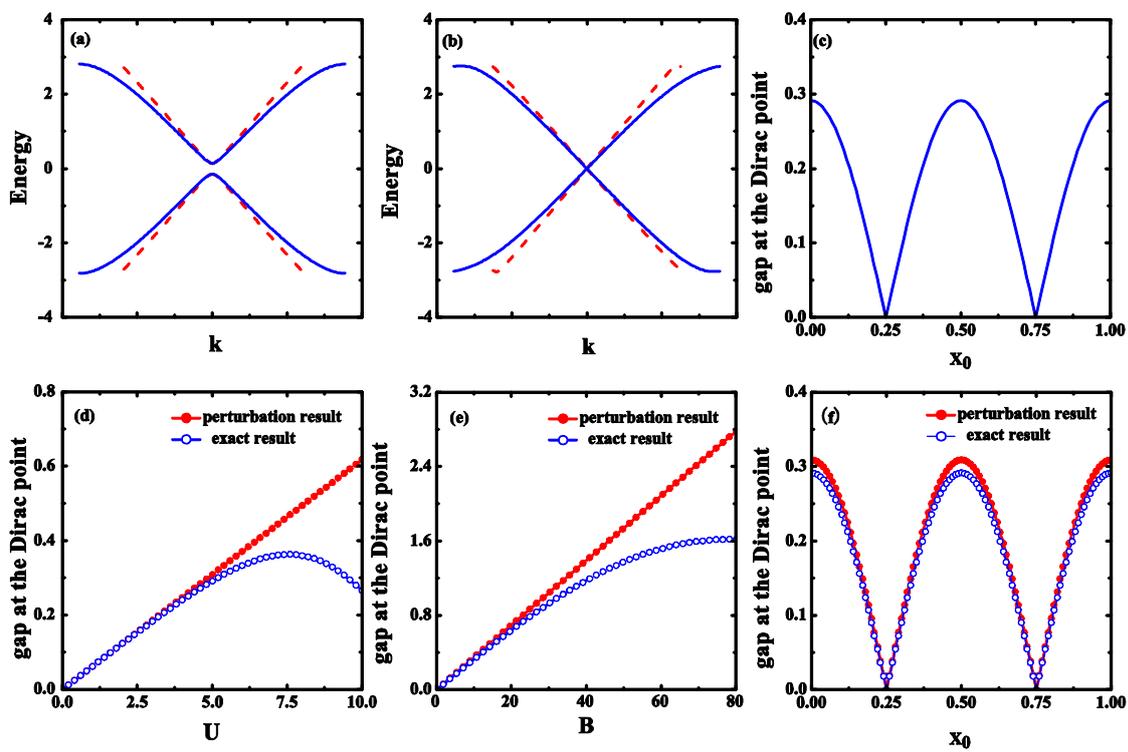

*Fig.2 by Xin Lin et al.*

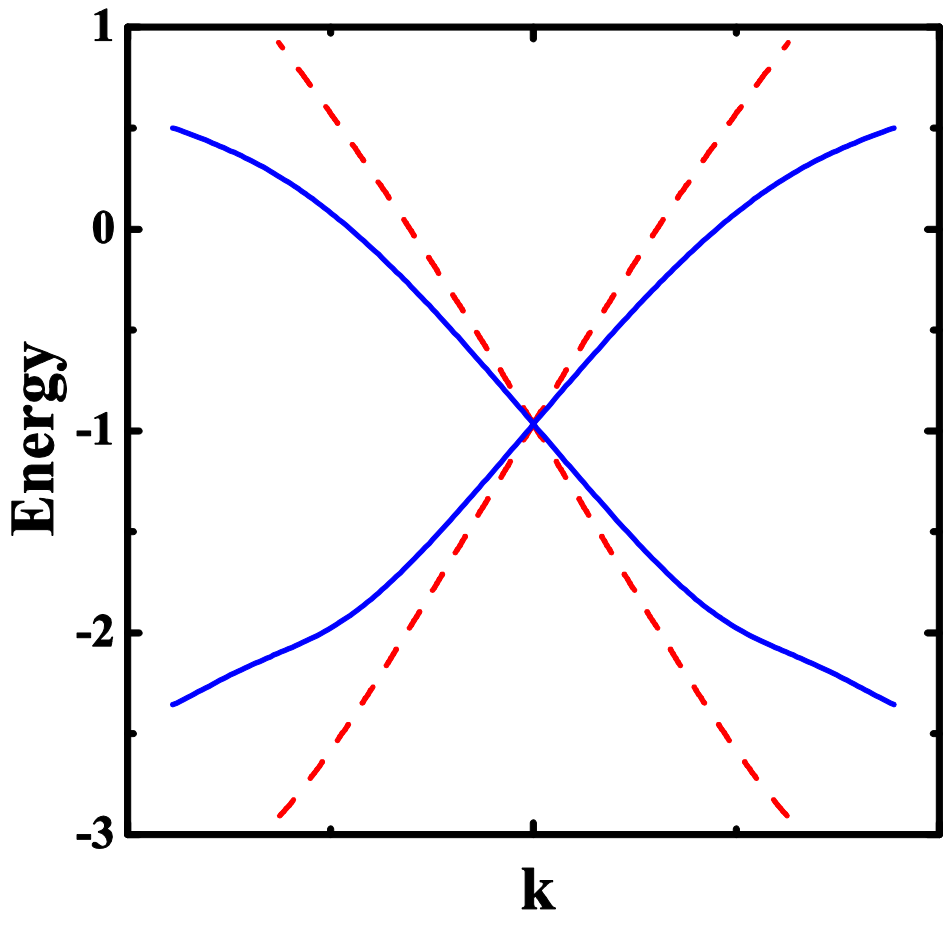

*Fig.3 by Xin Lin et al.*

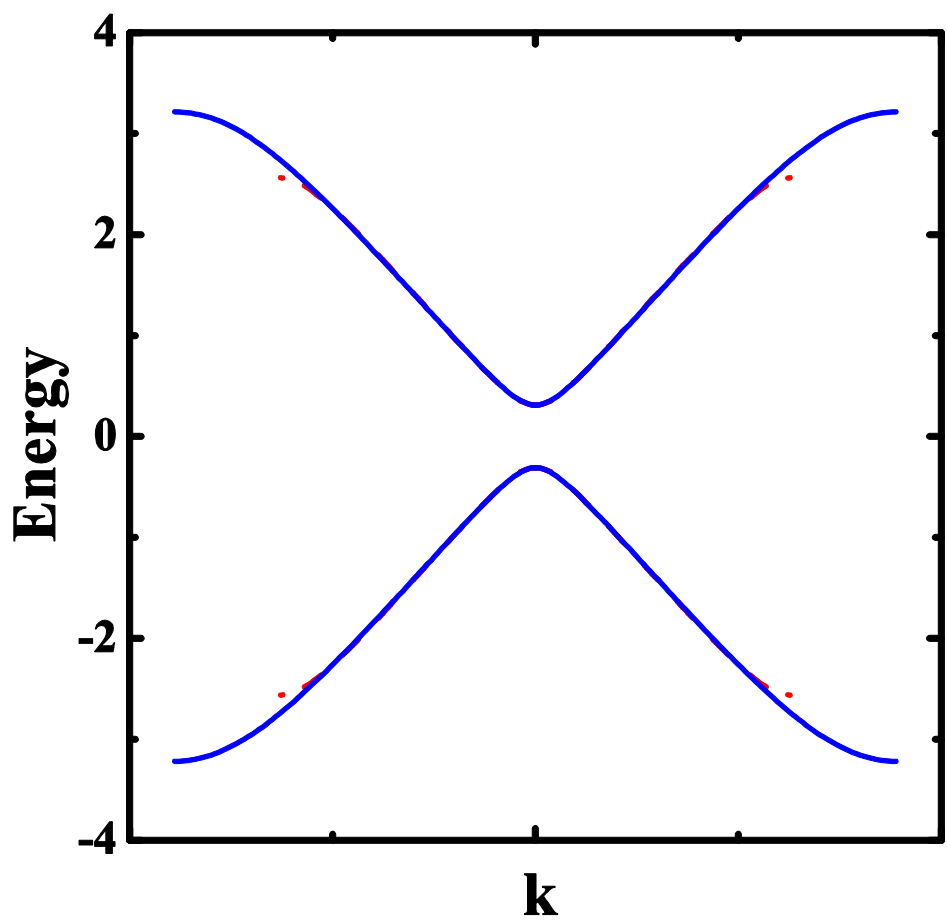

*Fig.4 by Xin Lin et al.*